\documentclass[12pt]{article}
\usepackage[T1]{fontenc}
\usepackage[latin2]{inputenc}
\usepackage{graphicx}

\title{
Algebraic Solution of RPA Equations for CC Quasi-elastic
Neutrino-Nucleus  Scattering}

\author{Krzysztof M. Graczyk\footnote{kgraczyk@ift.uni.wroc.pl},
        Jan T. Sobczyk\footnote{jsobczyk@ift.uni.wroc.pl}}

\begin{document}
\rightline{March 2003 IFT UWr 0320/2003, revised}
\begin{center}
\LARGE Algebraic solution of RPA equations for CC quasi-elastic
neutrino-nucleus  scattering
\\[4mm]
\end{center}

\begin{center}
\large
Krzysztof M. Graczyk\footnote{kgraczyk@ift.uni.wroc.pl},
        Jan T. Sobczyk\footnote{jsobczyk@ift.uni.wroc.pl}
\\[2mm]
July 15, 2003
\end{center}

\center{ Institute of Theoretical Physics,
University of Wroc\l aw              \\
pl. M. Borna 9, 50 - 204 Wroc\l aw, Poland}

\abstract{ Algebraic solution of RPA equations for nucleon
re-interactions in the case of quasi-elastic charged current
neutrino-nucleus scattering is presented. Abelian algebra of
matrices allows to extract four independent corrections to cross
section separately. Results of numerical computations are shown.}
\section{Introduction}

A better theoretical understanding of nuclear effects in
neutrino-nucleus scattering is important in view of data analysis
from new more precise neutrino experiments such as K2K, MINOS,
MiniBoone. Determination of parameters governing  neutrino
oscillation phenomenon, in particular of $\theta_{13}$ requires an
improved knowledge of neutrino-nucleus cross sections
\cite{ogolna}. In the above mentioned experiments neutrino beam
energy is of the order of 1 GeV. A characteristic feature of
neutrino-nucleus reactions in this energy domain is formation of
resonances and subsequent pion production. However, the
quasi-elastic contribution is still important and its precise
determination is of interest.

Nuclear effects in neutrino-nucleus interaction are often
evaluated in a framework of Monte Carlo approach. Scattering is
split into two steps. Neutrino interacts first with a free
nucleon, outgoing particles are then subject to re-interactions
inside nucleus. In more systematic theoretical approaches Mean
Field Theory with relativistic Fermi gas of protons and neutrons
as a ground state can serve as one of techniques to provide a
model for nucleus \cite{Walecka}. In its simplest version nucleus
forms a sea of fermions with momentum uniformly distributed inside
the Fermi sphere. The effect of MFT is that one has to substitute
nucleon mass $M$ by an effective mass $M^*$. It is well known that
for energies in the GeV range in the case of electron-nucleus
scattering Fermi gas model with fine tuned values of Fermi
momentum  and effective mass accounts for basic features of the
dynamics  \cite{Smith_Moniz}. More realistically, nucleons
interact with each other exchanging pions and $\rho$ mesons and
also short range correlations have to be considered by introducing
suitable contact interactions terms \cite{int}. In the ring
approximation ot the RPA approach a summation over all Feynman
diagrams is substituted by a sum of diagrams, where only 1p-1h
(one particle - one hole) excitation are included \cite{Fetter}.
In order to make the theory better one should also consider
elementary 2p-2h excitations in order to enlarge cross section in
the so-called "dip" region \cite{Meziani,Marteau}. It is however
difficult to include this contribution in the RPA scheme
\cite{Marteau_communikacja}.

In this paper analytic expressions for four contributions to RPA
corrections are derived in the case of quasi-elastic neutrino
reactions. An algebra of matrices is introduced to solve Dyson
equation. Results for separate contributions are presented. It is
known \cite{Langancke} that RPA corrections typically reduce the
maximum in the energy transfer differential cross section by a
factor of about 10\%. Our formalism when applied to the CC process
is only expected to reproduce these results. In particular we want
to mention here the paper \cite{Horowitz1}. We try to keep the
same notation in order to make comparison easier. A first
motivation for the present study is to construct a general
framework in which more detailed analysis of quasi-elastic CC
processes could be possible. One can evaluate significance of
uncertainties in various parameters: nucleons form factors,
coupling constants, effective mass etc. The second motivation is
that the same algebraic framework can be applied to NC reactions
and hopefully also (with necessary modifications) $\Delta$
excitation. In our RPA computations we keep a constant value of
Fermi momentum. Also $M^*$ is assumed to be a function of Fermi
momentum only thus it is a constant. Inclusion of local density
effects in the analytical framework is in principle possible but
rather complicated - a lot of manipulations with spherical
harmonics are necessary \cite{Marteau} with approximations
difficult to control. For all practical purposes it is sufficient
to have exact cross section formula for a fixed value of Fermi
momentum (and $M^*$) since one can perform a numerical integration
over Fermi momenta with a distribution defined by a density
profile of a nucleus in question. Our algebraic scheme enable
rather simple computation of both effects (RPA and local density)
and this is the third motivation for our work. We wish to mention
that similar idea of solving Dyson equation can be found in
\cite{Mornas}.

The paper is organized as follows. A short description of the
model, Feynman rules and the main technical tool - algebra of
matrices is given in section 2. Explicit expressions for RPA
corrections are presented in section 3. Section 4 contains a
discussion of results and also a comparison with RPA computations
done in the relativistic generalization of the Marteau model
\cite{Sobczyk}. There is a very nice agreement between two
approaches. Some technical details of algebraic computations are
collected in Appendix. The aim of this paper is to present main
features of the approach. More details and discussion will be
contained in the next paper being in preparation.

\section{The formalism }
\begin{figure}[ht]
\centerline{
\includegraphics[width=10.5cm]{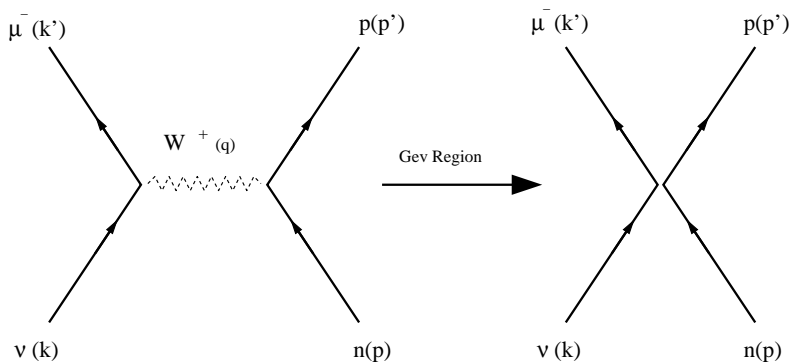}
} \caption{\footnotesize\footnotesize\label{neutrino} The basic
diagram describing neutrino-nucleus interaction.  Scattering takes
place on a single nucleon with a definite momentum given by a
Fermi gas distribution. In energy domain of few GeV effective
four-Fermion vertex provides an excellent approximation.}
\end{figure}

We consider charged-current (CC) quasi-elastic neutrino-nucleus
scattering (Fig. \ref{neutrino}).
The model of nucleus is given by the Mean Field Theory
\cite{Walecka} (nucleon mass becomes an effective mass $M^*(k_F)$)
with interactions due to contact terms and exchange of pions and
$\rho$ mesons \cite{int}. In a first approximation the nucleus is
treated as a relativistic Fermi gas with Fermi momentum $k_F$
determined by nucleons density.

Elementary weak charge current nucleon-nucleon current is
expressed by means of form factors \cite{LS}:
\begin{equation}
\Gamma^\alpha(q_\mu) = F_1(q^2_\mu)\gamma^\alpha
+F_2(q_\mu^2)\frac{\mathrm{i}\sigma^{\alpha\nu}q_\nu}{2M}
+G_A(q_\mu^2)\gamma^\alpha\gamma^5
\end{equation}
We omit $G_p(q_\mu)$ term since its contribution to $\nu_{e}$ and
$\nu_{\mu}$ cross sections at $E_{\nu}\sim 1GeV$ is negligible. It
can be put into our scheme if required with only minor modifications.

The differential cross section (per nucleon) reads:
\begin{equation}
\frac{d^2 \sigma }{d|\vec{q}| \, dq_0}=
-\frac{ G_F^2 \mathrm{cos}^2\theta_c \; |\vec{q}|}{16 \pi^2 \rho_F E^2 }
\mathrm{Im}
\left (
{L_\mu}^\nu {\Pi^\mu}_\nu
\right ).
\end{equation}
$\rho_F = k_F^3/3\pi^2$, we assume the same values of Fermi
momentum for neutrons and protons. $L^{\mu\nu}$ is a
leptonic tensor:
\begin{equation}
L_{\mu\nu} = 8 \left ( k_\mu {k'}_\nu + {k'}_\mu k_\nu -
g_{\mu\nu}k_\alpha{k'}^\alpha \pm
\mathrm{i}\epsilon_{\mu\nu\alpha\beta}{k'}^\alpha k^\beta \right
).
\end{equation}

The sign $\pm$ depends on a process considered
(neutrino/antineutrino).

The polarization tensor $\Pi^{\mu\nu}$ is a basic object that
contains full information about nuclear effects. It is defined as
a chronological product of many body currents:
\begin{equation}
\Pi^{\mu\nu}(q_0, q) = -\mathrm{i}\int d^4 x e^{\mathrm{i}
q_\alpha x^\alpha} <0|T\left (\mathcal{J}^\mu(x)\mathcal{J}^\nu(0)
\right ) |0> .
\end{equation}

The $|0>$ is a ground state of nucleus described by the Fermi gas
model. Prescription for nuclear physics \cite{Walecka_old} enables
evaluation of $\Pi^{\mu\nu}$ by means of the standard QFT
techniques with modified (depending on $k_F$) progagator $G(p)$.

\subsection{ Feynman rules }

The polarization tensor is split into a "free" part and RPA
correction:
\begin{equation}
\Pi^{\mu\nu} = \Pi^{\mu\nu}_{free} + \Delta\Pi_{RPA}^{\mu\nu} .
\end{equation}
The "free" tensor is given by a simple fermion loop
\begin{figure}[h]
\centerline{
\includegraphics[width=2cm]{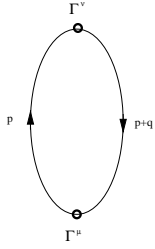}
} \caption{\footnotesize\footnotesize\label{free}
$\Pi_{free}^{\mu\nu}$ }
\end{figure}
(Fig. \ref{free}) which is spanned between two vertices with form
factors insertions.
\begin{equation}
\Pi_{free}^{\mu\nu}(q)  = -\mathrm{i} \int \frac{d^4 p}{( 2 \pi
)^4} \mathrm{Tr} \left ( G(p + q)\Gamma^\mu(q)G(p)\Gamma^\nu (-q )
\right )
\end{equation}
\begin{equation}
 G(p) = (p \!\!\!\! / + M^* ) \left ( \frac{ 1 }{
p_\alpha^2 - {M^*}^2 + \mathrm{i}\epsilon } +
\frac{\mathrm{i}\pi}{E_p}\delta(p_0 - E_p)\theta(k_F - p) \right
).
\end{equation}
$G(p)$ describes a propagation of a free fermion in the Fermi sea.

RPA corrections arise from summation of an infinite sum of 1p-1h
contributions \cite{Fetter}. These corrections are given by
Feynman diagrams containing $\pi$ propagator - $V^{\mu\nu}$,
$\rho$ propagator - $W^{\mu\nu}$ and interactions vertices $NN\pi$
as well as $NN\rho$. Landau-Migdal parameter $g'$ is put together
with genuine pion propagator to form a redefined pion "propagator"
(for details see \cite{Horowitz1}).

In what follows in this section we do not write down explicitly
Lorentz indices. It will be understood that unless specified all
the objects are $4 \times 4$ matrices with indices
${(\;\;)_\mu}^\nu$.
\begin{figure}
\centerline{ \includegraphics[width=9cm]{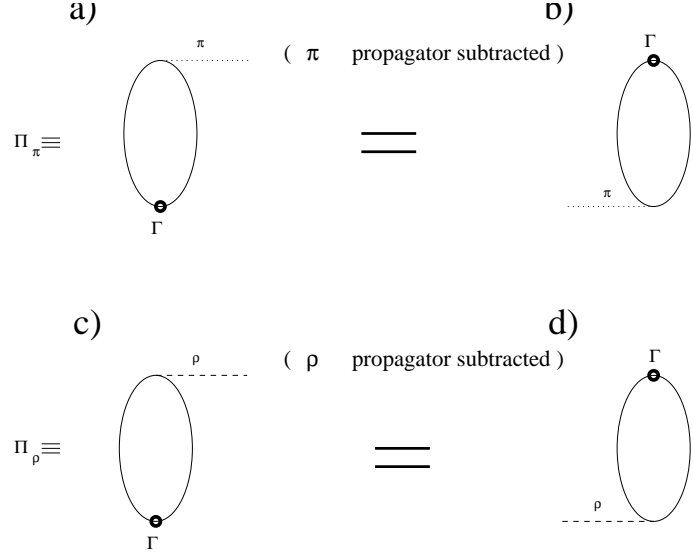} } \caption{
\footnotesize \label{rho_pi} Diagrammatical definitions of
$\Pi_{\pi ,\rho}$. They are loops with CC weak nucleon-nucleon
current in one vertex and $NN\pi$ or $NN\rho$ vertices. }
\end{figure}
We define tensors $\Pi_\rho$ and $\Pi_\pi$ as loop diagrams with
CC weak nucleon-nucleon current in one vertex and $NN\rho$ or
$NN\pi$ vertices (Fig.\ref{rho_pi}). Notice that contributions
from tensors given by figures (\ref{rho_pi}a) (\ref{rho_pi}b) are
equal. The same applies to contributions (\ref{rho_pi}c) and
(\ref{rho_pi}d). This property simplifies the algebraic form of
RPA corrections.
\begin{figure}
\centerline{
\includegraphics[width=8cm]{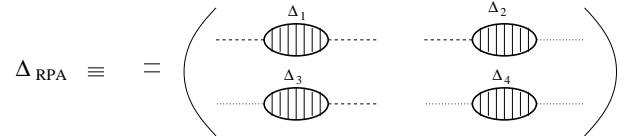}
} \caption{\footnotesize\footnotesize\label{propagator_rpa}
Following \cite{Horowitz1} we define RPA propagator as a sum over
1p-1h diagrams with external meson fields propagators.}
\end{figure}

\begin{figure}
\centerline{
\includegraphics[width=9cm]{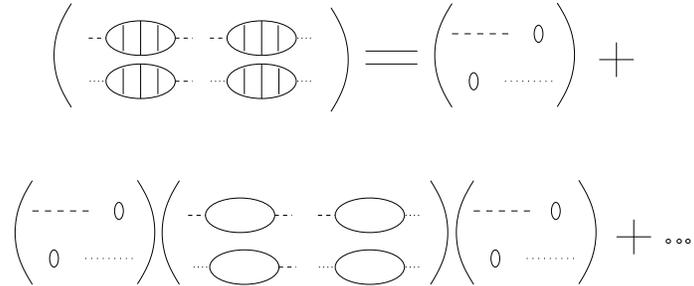}
} \caption{\footnotesize\footnotesize\label{szereg} The propagator
RPA can be expressed as an infinite sum.}
\end{figure}

$\Delta_{RPA}$ is an $8 \times 8$  matrix  defined by an infinite series:
\begin{equation}
\label{suma} \Delta_{RPA} = \Delta_0
+ \Delta_0 \Pi_G \Delta_0 +
\Delta_0 \Pi_G \Delta_0 \Pi_G \Delta_0 + ... ,\end{equation} which
is illustrated in Fig. \ref{szereg}.
\begin{figure}
\centerline{\includegraphics[width=6cm]{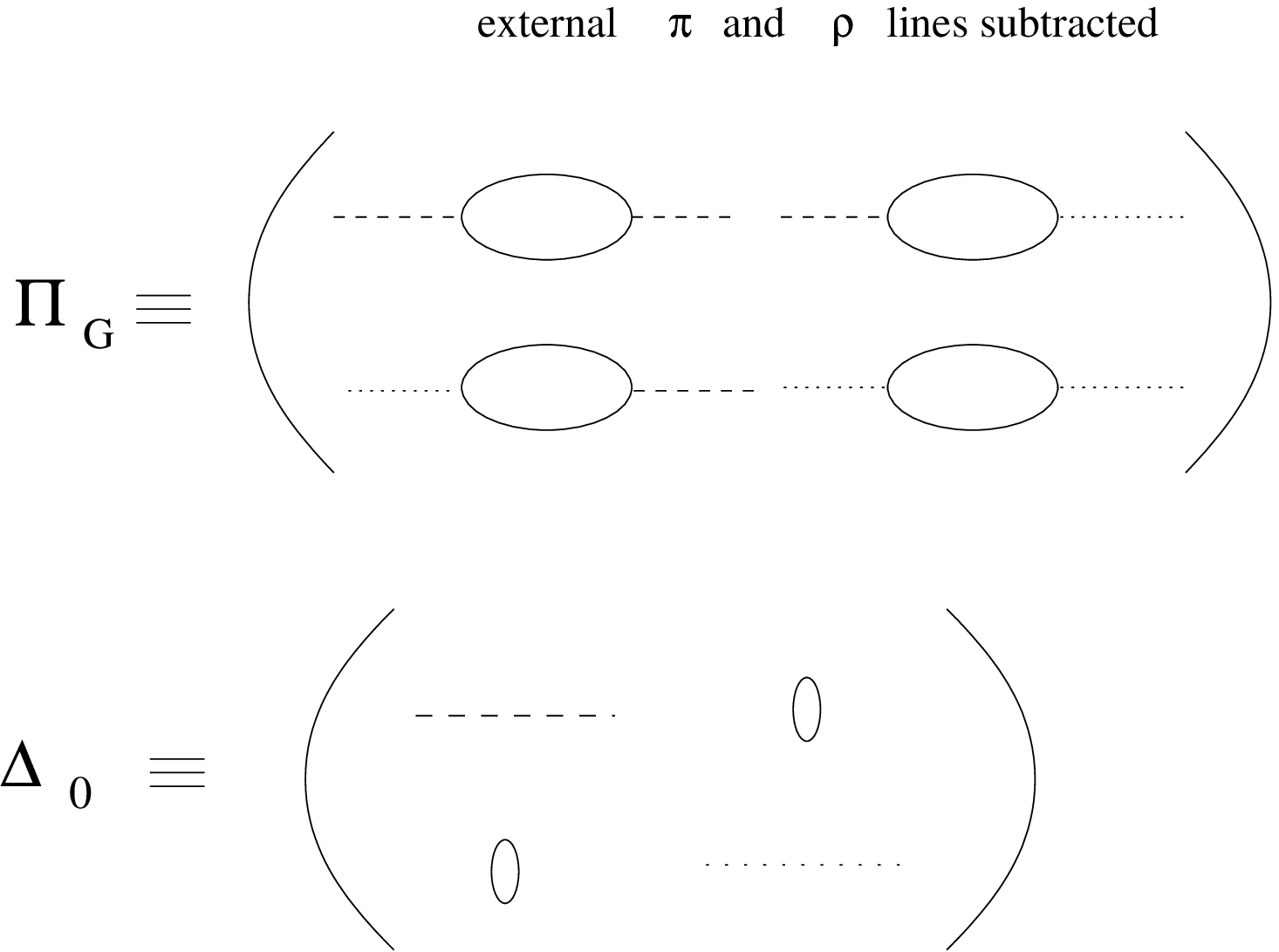}}
\caption{\footnotesize\footnotesize\label{generic_D0} Tensors
$\Pi_G$ and $D_0$. }
\end{figure}
We defined two new tensors - $8 \times 8$ matrices: $\Pi_G$ and
$\Delta_0$. Their definitions can be simply understood in
diagrammatical language (Fig. \ref{generic_D0}).

\begin{figure}
\centerline{
\includegraphics[width=10cm]{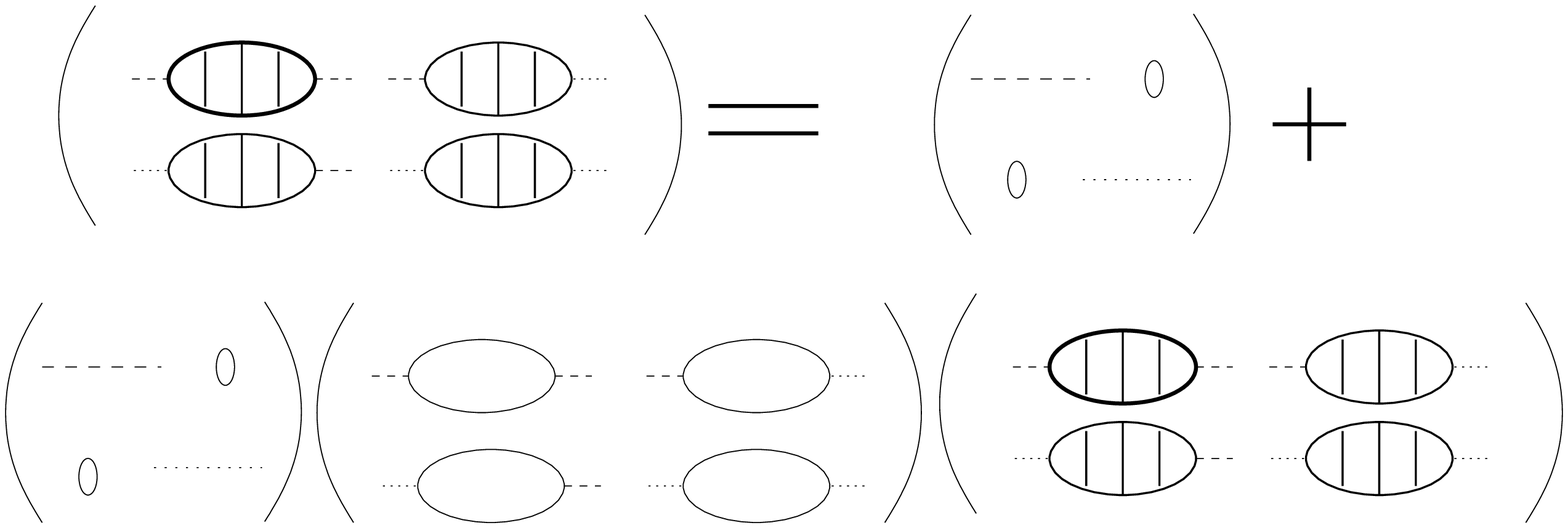}
} \caption{\footnotesize\label{rownaniedyson} Dyson equation for
RPA propagator.}
\end{figure}
With all these definitions the formula (\ref{suma}) can be
rewritten in the form of Dyson equation (Fig.
\ref{rownaniedyson}):

\begin{equation}
\label{Dyson}
\Delta_{RPA} = \Delta_0 +
\Delta_0 \Pi_G \Delta_{RPA}
.\end{equation}

\begin{figure}
\centerline{
\includegraphics[width=9cm]{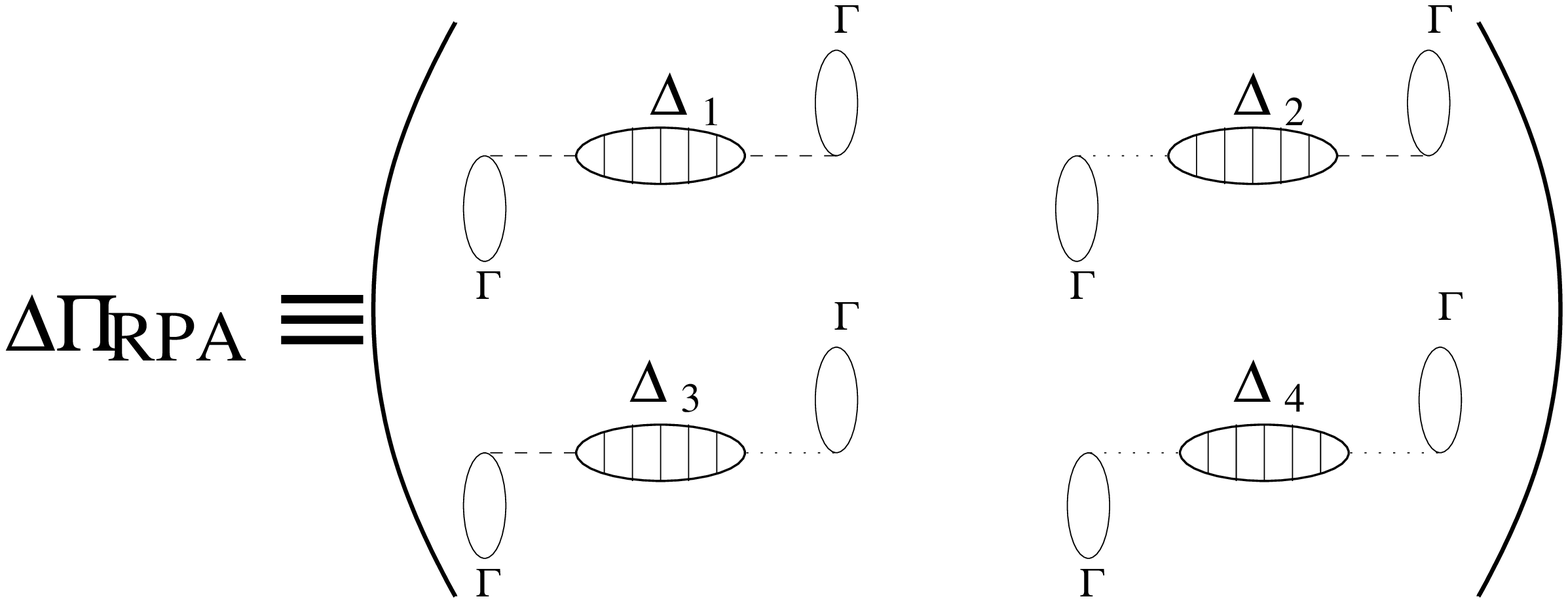}
} \caption{\footnotesize\footnotesize \label{deltarpa}
Diagrammatical explanation of RPA corrections to polarization
propagator.}
\end{figure}

It is clear that RPA corrections to polarization propagator are
given by $\Delta_{RPA}$ multiplied from both sides by $\Pi_\rho$
and $\Pi_\pi$
\begin{equation}\label{kkk}
\Delta \Pi_{RPA}(q_\mu) = \pmatrix{ \Pi_\rho(q_\mu) &
\Pi_\pi(q_\mu)} \Delta_{RPA}(q_\mu) \pmatrix{ \Pi_\rho(q_\mu) \cr
\Pi_\pi(q_\mu)}.
\end{equation}

The diagrammatic explanation of this formula is presented in {Fig.
\ref{deltarpa}}. Our strategy is to solve equation for
$\Delta_{RPA}$ and then to obtain expression for
$\Delta\Pi_{RPA}$.

\subsection{ Algebraic properties of the polarization tensor }

A coordinate system is chosen in which four momentum transfer
reads:
\begin{equation}
q_\mu = k_\mu - {k'}_\mu = (q_0, q, 0, 0).
\end{equation}
The Dyson equation is a $8 \times 8$ matrix equation. In order to
solve it we introduce $4 \times 4$ matrices:
\begin{eqnarray} e_L   =    \pmatrix{ -\frac{q^2}{q_\mu^2}  &
\frac{q_0 q}{q_\mu^2} & 0 & 0 \cr -\frac{q_0 q}{q_\mu^2}&
\frac{q_0^2}{q_\mu^2} & 0 & 0 \cr
          0 & 0 & 0 & 0  \cr
          0 & 0 & 0 & 0}
&  &
e_T = \pmatrix{ 0 & 0 & 0 & 0   \cr
                 0 & 0 & 0 & 0  \cr
                0 & 0 &  1 & 0  \cr
                0 & 0 &  0 & 1 }
\nonumber \\
e_A = \pmatrix{ 1 & 0 & 0 & 0 \cr
                0 & 1 & 0 & 0 \cr
                0 & 0 & 1 & 0 \cr
                0 & 0 & 0 & 1 }
& &
e_{VA} = \pmatrix{
             0 & 0 & 0 & 0 \cr
             0 & 0 & 0 & 0 \cr
             0 & 0 & 0 & -i \cr
             0 & 0 & i & 0
                 }
\end{eqnarray}
They satisfy matrix multiplication relations:
\begin{eqnarray}
e_L \, e_L       & = & e_L                    \nonumber \\
e_L \, e_T       & = & 0  \;\;\; = \;\; e_T\, e_L \nonumber \\
e_L \, e_A       & = & e_L \,\; = \;\; e_A\, e_L \nonumber \\
e_L \, e_{VA}    & = & 0   \;\;\; = \;\; e_{VA}\, e_L  \nonumber \\
e_T \, e_T       & = & e_T                    \nonumber \\
e_T \, e_A       & = & e_T \,\; = \;\; e_A\, e_T \nonumber \\
e_T \, e_{VA}    & = & e_{VA}\; = \;\; e_{VA} \, e_T \nonumber \\
e_A \, e_A       & = & e_A                    \nonumber \\
e_A \, e_{VA}    & = & e_{VA} \; = \;\; e_{VA}\, e_A       \nonumber  \\
e_{VA} \; e_{VA} & = & e_T .\end{eqnarray}

The polarization tensor must be a linear combinations of $e_L$,
$e_T$, $e_A$, $e_{VA}$,
\begin{equation}
\Pi = \Pi^L e_L +\Pi^T e_T + \Pi^{VA} e_{VA} + \Pi^A e_A .
\end{equation}

This is clear since $\Pi$ is given by a sum over Feynman diagrams
and each contribution is of this form due to the fact that algebra
$e_L, e_T, e_{VA}, e_A$ is closed under multiplication and all the
building blocks are expressed in terms of these four basic
matrices. Consequently the cross section has the form
\begin{equation}
\frac{d^2 \sigma }{dq \, dq_0}  = -\frac{G_F^2
\mathrm{cos}^2\theta_c \; q}{16 \pi^2 \rho_F E^2 } \mathrm{Im}
\left (L_L \Pi^L + L_T \Pi^T \pm L_{VA} \Pi^{VA} + L_A \Pi^A\right
), \end{equation} where $L_L \equiv  {L_\mu}^\nu {{e_L}^\mu}_\nu$
etc. Many authors, e.g. \cite{Horowitz1} call four contributions:
longitudinal, transverse, V-A, and axial. This can cause some
confusion because contributions to the cross section are sometimes
called after spin-isospin operators present in nucleon-nucleon
transition current. We will return to this point in Section 4.

\section{ RPA corrections }
A lot of simplifications come from the fact that $\Delta_0$
contains only longitudinal, transverse and axial terms
\cite{Horowitz1}.
\begin{equation}
\Delta_0   = \pmatrix{ W & 0 \cr
          0 & V   }
=\pmatrix{ W^L e_L + W^T e_T + W^A e_A & 0 \cr
                      0 &  V^L e_L + V^T e_T + V^A e_A}
.\end{equation}

$W$ and $V$ are

\begin{eqnarray}
W^L = \; W^T & = & - \frac{q_\mu^2/m_\rho^2}{q_\mu^2 - m_\rho^2 + \mathrm{i}\epsilon}\\
W^A & = &  \frac{q_\mu^2 - m_\rho^2}{m_\rho^2}
\frac{1}{q_\mu^2 - m_\rho^2 + \mathrm{i}\epsilon}\\
V^L  = \; V^T & = & - \frac{q_\mu^2}{q_\mu^2 - m_\pi^2 + \mathrm{i}\epsilon}\\
V^A & = &   \frac{q_\mu^2}{q_\mu^2 - m_\pi^2 + \mathrm{i}\epsilon} -g'\\
\end{eqnarray}
$g'$ is the Landau Migdal parameter.

Also
\begin{eqnarray}
\Pi_\rho & = &
\Pi_\rho^L e_L +\Pi_\rho^T e_T + \Pi^{VA}_\rho e_{VA} \\
\Pi_\pi & = &
\Pi^L_\pi e_L + \Pi_\pi^T e_T + \Pi^{VA}_\pi e_{VA} + \Pi_\pi^A e_A
\end{eqnarray}

We find
\begin{eqnarray}
\Pi_G  = \pmatrix{\Pi_{\rho\rho} & \Pi_{\rho\pi} \cr
         \Pi_{\pi\rho}  & \Pi_{\pi\pi} }=
\pmatrix{ \Pi_{\rho\rho}^L e_L + \Pi_{\rho\rho}^T e_T &
\Pi_{\rho\pi}^{VA} e_{VA}  \cr
\Pi_{\rho\pi}^{VA} e_{VA} &
 \Pi_{\pi\pi}^L e_L + \Pi_{\rho\rho}^T e_T  + \Pi_{\rho\rho}^A e_A}
\end{eqnarray}
We introduce a general notation:
\begin{equation}
\Delta_{RPA}  =  \pmatrix{ \Delta_1 & \Delta_2 \cr \Delta_3 &
\Delta_4 }
\end{equation}
$$
\Delta_i  =  \Delta_i^{L}e_L + \Delta_i^{T}e_T +
\Delta_i^{VA}e_{VA} + \Delta_i^{A}e_A \;\;\; i = 1,2,3,4.
$$
The $8 \times 8$ matrix equation (\ref{Dyson}) can be rewritten as
a set of  four $4 \times 4$ matrix equations:
\begin{eqnarray}
\label{pierwsze}
\Delta_1 & = & W + W\Pi_{\rho\rho}\Delta_1  + W\Pi_{\rho\pi}\Delta_3 \nonumber \\
\label{drugie}
\Delta_2 & = & W\Pi_{\rho\rho}\Delta_2 + W \Pi_{\rho\pi}\Delta_4 \nonumber \\
\label{trzecie} \Delta_3 & = & V\Pi_{\rho\pi}\Delta_1  +
V \Pi_{\pi\pi}\Delta_3 \nonumber \\
\label{czwarte} \Delta_4 & = & V + V\Pi_{\rho\pi}\Delta_2 + V
\Pi_{\pi\pi}\Delta_4 .
\end{eqnarray}

Equations (\ref{czwarte}) are transformed to a set of algebraic
equations and solved in the Appendix. We obtain a general solution
for $\Delta_{RPA}$
\begin{equation}
\Delta_{RPA} = \pmatrix{ \Delta^A_1 e_A + \Delta_1^L e_L
+\Delta_1^T e_T & \Delta^{VA}_2 e_{VA} \cr \Delta_3^{VA} e_{VA} &
\Delta^A_4 e_A + \Delta_4^L e_L + \Delta_4^T e_T} .
\end{equation}

RPA corrections to polarization tensor are:
\begin{eqnarray}
\label{poprawka_l}
\Delta \Pi_{RPA}^L & = &
\left (
(\Pi_\pi^L)^2 + 2\Pi_\pi^L \Pi^A_\pi
\right )
\left (
\Delta_4^A + \Delta^L_4
\right ) \nonumber\\
& + &  (\Pi_\pi^A)^2 \Delta^L_4
+(\Pi_\rho^L)^2 \left ( \Delta_1^A + \Delta_1^L\right)\\
\label{poprawka_t}
\Delta \Pi_{RPA}^T & = &
\left(
 (\Pi_\pi^T)^2 + 2\Pi_\pi^T \Pi^A_\pi + (\Pi_\pi^{VA})^2
\right )
\left (
\Delta_4^A + \Delta^T_4
\right )
+ (\Pi_\pi^A)^2 \Delta^T_4 \nonumber \\
& + &
\left (
(\Pi_\rho^T)^2 + (\Pi_\rho^{VA})^2
\right )
\left(
\Delta_1^T +\Delta_1^A
\right)\nonumber \\
& + &
\left (
\Pi_\rho^{VA}\Pi_\pi^T + \Pi_\rho^{VA} \Pi_\pi^A
+ \Pi_\rho^T \Pi_\pi^{VA}
\right )\left( \Delta_2^{VA} + \Delta_3^{VA}\right )\\
\label{poprawka_va}
\Delta \Pi_{RPA}^{VA} & = &
\left (
\Pi_\rho^T\Pi_\pi^T  + \Pi_\rho^T\Pi_\pi^A
+ \Pi_\rho^{VA} \Pi_\pi^{VA}
\right )
\left( \Delta_2^{VA} + \Delta_3^{VA}\right )\\
& &\!\!\!\!\!\!\!\!\!\!\!\!\!\!\!\!\!\!\!\!\!\!\!\!
+2 \left (\Delta_1^T + \Delta_1^A \right )\Pi_\rho^{VA}\Pi_\rho^T +
2\left (\Delta_4^T + \Delta_4^A \right )
\left ( \Pi_\pi^A + \Pi_\pi^T\right )\Pi_\pi^{VA} \\
\label{poprawka_a}
\Delta \Pi_{RPA}^A & = & (\Pi_\pi^A)^2 \Delta_4^A
\end{eqnarray}

Expressions for $\Delta_1^A$... are written in the appendix.

\section{Discussion}

\begin{figure}[p]
\centerline{
\includegraphics[width=10cm]{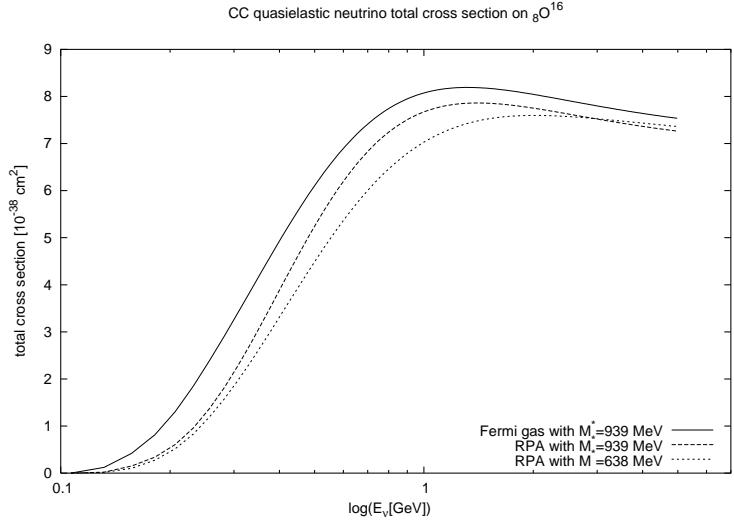}
} \caption{\footnotesize\footnotesize Comparison of the computations of quasielastic neutrino total cross
sections on oxygen,  $g'=0.7$.
\label{rys6}}
\end{figure}
\begin{figure}[p]
\centerline{
\includegraphics[width=10cm]{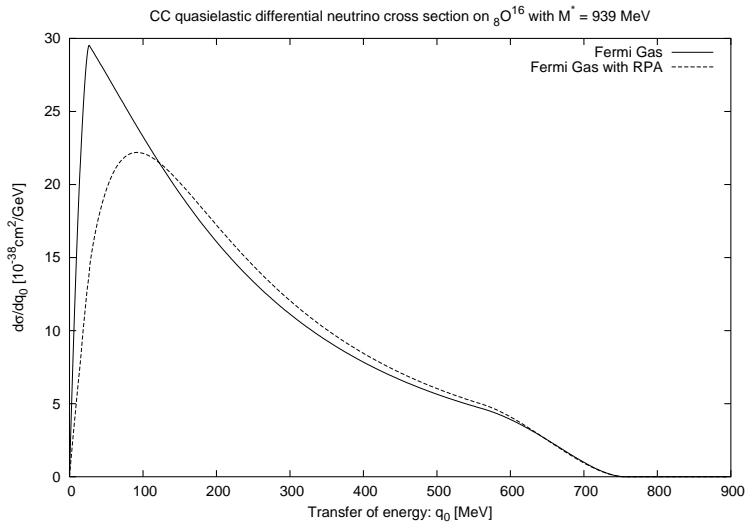}
} \caption{\footnotesize\footnotesize Comparison of differential quasielastic neutrino
cross section for Fermi Gas of free nucleons with cross section
modified by RPA corrections for neutrino energy 1 GeV. The
calculation was done for the free nucleon mass and $g'=0.7$.
\label{rys1} }
\end{figure}
\begin{figure}[p]
\centerline{
\includegraphics[width=10cm]{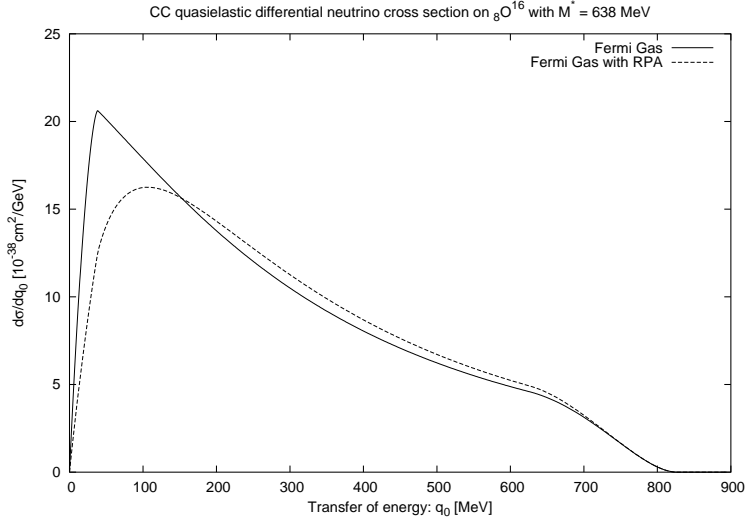}
} \caption{\footnotesize\footnotesize Comparison of differential quasielastic neutrino
cross section for Fermi Gas of free nucleons with cross section
modified by RPA corrections for neutrino energy 1 GeV.
Calculations were done with effective mass $M^* = 638$ MeV,
$g'=0.7$. \label{rys2} }
\end{figure}
\begin{figure}[p]
\centerline{
\includegraphics[width=10cm]{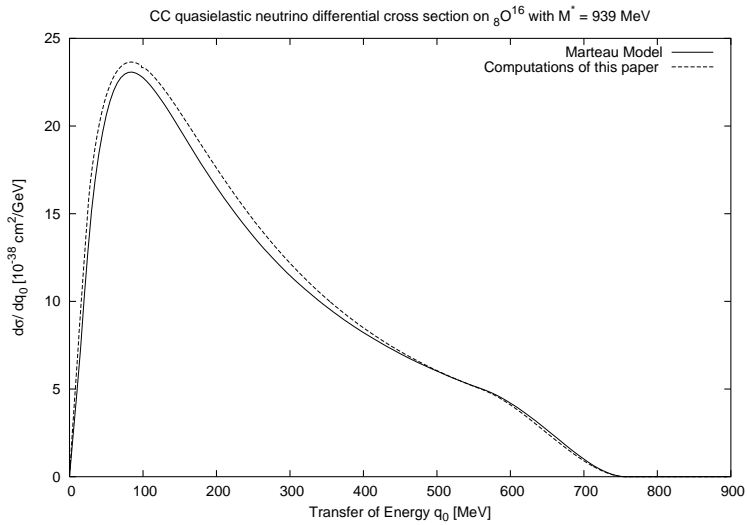}
} \caption{\footnotesize\footnotesize Comparison of differential quasielastic neutrino
cross sections with RPA corrections in: the model of this paper
and Marteau model\cite{Sobczyk} for neutrino
energy 1 GeV. The calculation was done for the free nucleon mass
and $g'=0.6$. \label{rys3} }
\end{figure}

\begin{figure}
\centerline{
\includegraphics[width=10cm]{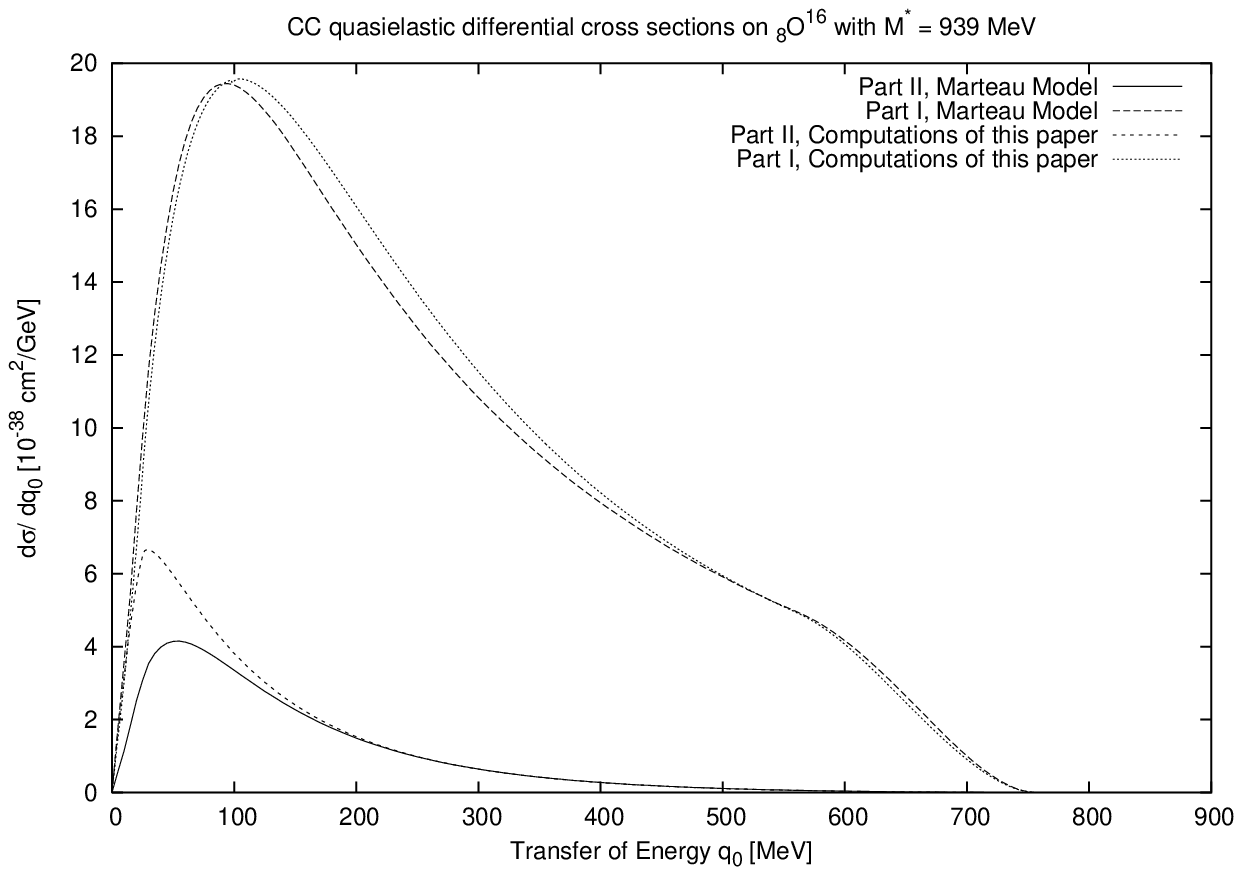}
} \caption{\footnotesize\footnotesize Comparison of RPA
corrections in contributions \textbf{I} and \textbf{II}, in
the model of this paper and Marteau model \cite{Sobczyk}.
The calculation was done for the free nucleon mass, $g'=0.6$.
\label{rys4} }
\end{figure}

Numerical evaluation of RPA corrections to $\Pi^{\mu\nu}$ requires
a knowledge of both real and imaginary parts of all objects
presented in (\ref{poprawka_l} - \ref{poprawka_a}). We get the
necessary formulae from \cite{tensory}. With this input one can
perform numerical analysis of our algebraic results.

We present some plots demonstrating consistence of our procedures.
We compare RPA corrections obtained in our paper with those
computed in a different approach. The approach we choose is based
on Marteau model with some improvements \cite{Sobczyk}. A use of
relativistic generalization of the Lindhard function made
kinematical regions of both models the same.

In the Marteau model three contributions to cross section are
identified according to spin-isospin operators present in the
transition amplitude. To find a bridge between two decompositions
we observe that  hadronic tensor components $H^{00}, H^{01},
H^{10}, H^{11}$ contribute only to longitudinal and charge
contributions (in spin-isospin nomenclature) while remaining
components contribute only to the transverse part. This is
strictly speaking true in the approximation when ${|\vec p|\over
M}$ ($\vec p$ is target nucleon momentum) terms are neglected in
$H^{\mu\nu}$ which is valid within few $\%$. We decided therefore
to single out two contributions in both approaches and to call
them in order to avoid confusion: {\bf I} and {\bf II}. The
contribution {\bf II} is equivalent to the sum of charge and
longitudinal parts in the Marteau approach while the contribution
{\bf I} is equivalent to the transverse part.

The identification of {\bf I} and {\bf II} parts in our approach
requires some algebra. We obtain:

\begin{eqnarray}
\left ( L_{\mu\nu}\Pi^{\mu\nu}\right)_{\bf I} &= L_T\left( \Pi^T -
\Pi^A \right ) + L_{VA} \Pi^{VA}\\
\left( L_{\mu\nu}\Pi^{\mu\nu}\right)_{\bf II} &= L_L \Pi^L + \left
(L_A + L_T \right)\Pi^A
\end{eqnarray}
In numerical calculations we assumed the following values of
parameters present in the theory: $M_A=1.03$ GeV - axial mass,
standard values of coupling constants for pions and $\rho$ mesons,
Landau Migdal parameter $g'=0.7$ except from two comparison plots
(in \cite{Sobczyk} the value $g'=0.6$ was assumed and we take the
same value in order to make the comparison consistent). Effective
mass was calculated according to self-consistency equation of MFT
theory \cite{Walecka}. It is assumed that target nucleus is oxygen
$^{16}O$ and that Fermi momentum is $k_F=225$ MeV. We get
$M^*=638$~MeV.

In the Fig.\ref{rys6} we compare predictions for the total cross
section in three cases: (i) free Fermi gas with $M^*=939$ MeV,
(ii) RPA computations with $M^*=939$ MeV, (iii) RPA computations
with $M^*=638$ MeV. Inclusion of RPA correlations makes the cross
section smaller. In the third case reduction of the cross section
is more significant for neutrino energies up to about $3$ GeV.

In the Fig.\ref{rys1} and Fig.\ref{rys2} we show differential
cross sections in energy transfer for neutrino  energy of $1$ GeV.
As above we distinguish two cases in which effective mass is taken
either as a free mass of nucleon or as $638$ MeV. One can see the
typical expected behavior: in the RPA case quasi-elastic peak
becomes reduced but at larger values of energy transfer an effect
of RPA is to increase slightly the cross section. We notice that
due to effective mass kinematically allowed regions in energy
transfer are in two cases different.

In last two figures we compare our differential cross sections
with predictions of the model described in \cite{Sobczyk}. A good
agreement between influence of RPA corrections in two models is
seen. The contribution \textbf{I} is dominant in both cases.
Differences between them are small. The behavior of contribution
\textbf{II} in both cases is similar. Marteau model gives rise to
smaller contributions at energy transfer of $\sim 50$ MeV and the
whole contribution becomes reduced by about $25 \%$.

We conclude that our algebraic solution of RPA equations leads to
modifications of cross section similar to other approaches. We
cannot expect that Marteau model \cite{Sobczyk} can produce
numerically identical results as it is a hybrid model which
combines nonrelativistic potential approach with relativistic
Lindhard function. We hope that our algebraic scheme will be
useful in other cases mentioned in the Introduction.

\section{Appendix A}
Equations (\ref{czwarte}) are rewritten as sixteen:
\begin{eqnarray}
\label{a}
\Delta_1^A & = & W^A \nonumber \\
\label{b}
\Delta_2^A & = & 0    \nonumber \\
\label{c}
\Delta_3^A & = & V^A \Delta_3^A \Pi_{\pi\pi}^A     \nonumber \\
\label{d} \Delta_4^A & = & V^A \left (1 + \Delta_4^A
\Pi_{\pi\pi}^A \right )
\end{eqnarray}
\begin{eqnarray}
\label{e} \Delta_1^L & = & R^L + \left ( \Delta_1^A +
\Delta_1^L\right )
\left ( W^A + W^L \right ) \Pi_{\rho\rho}^L    \nonumber \\
\label{f} \Delta_2^L & = & \left ( \Delta_2^A + \Delta_2^L \right)
\left ( W^A +  W^L \right ) \Pi_{\rho\rho}^L \nonumber \\
\Delta_3^L & = & \Delta_3^A \left ( \Pi_{\pi\pi}^A  V^L +
\Pi_{\pi\pi}^L V^A
+ \Pi_{\pi\pi}^L V^L \right ) \nonumber \\
\label{g} &  & + \Delta_3^L \left ( V^A +V^L  \right )
  \left (\Pi_{\pi\pi}^A  +  \Pi_{\pi\pi}^L \right ) \nonumber \\
\Delta_4^L & = & V^L +   \Delta_4^A \left ( \Pi_{\pi\pi}^A V^L +
\Pi_{\pi\pi}^L  V^A +
\Pi_{\pi\pi}^L V^L \right ) \nonumber\\
\label{h} &   & +\Delta_4^L \left (  V^A + V^L \right ) \left (
\Pi_{\pi\pi}^L + \Pi_{\pi\pi}^A  \right )
\end{eqnarray}
\begin{eqnarray}
\Delta_1^T & = & W^T + \left ( \Delta_1^A  + \Delta_1^T \right )
\Pi_{\rho\rho}^T \left ( W^A + W^T \right ) \nonumber\\
&  & \label{i}
+\Delta_3^{VA}\Pi_{\rho\pi}^{VA}\left ( W^A + W^T\right ) \nonumber \\
\label{j} \Delta_2^T & = & \left (\Delta_2^A + \Delta_2^T \right
)\Pi_{\rho\rho}^T \left (W^A + W^T \right )
+ \Delta_4^{VA} \Pi_{\rho\pi}^{VA}\left ( W^A + W^T \right )\nonumber \\
\Delta_3^T & = & \Delta_1^{VA} \Pi_{\rho\pi}^{VA}\left ( V^A + V^T
\right ) + \Delta_3^T \left ( \Pi_{\pi\pi}^A + \Pi_{\pi\pi}^T
\right )
\left ( V^A + V^T \right )\nonumber \\
\label{k} & & + \Delta_3^A \left ( \Pi_{\pi\pi}^A V^T +
\Pi_{\pi\pi}^T V^A + \Pi_{\pi\pi}^T V^T
\right ) \nonumber \\
\Delta_4^T & = &   V^T + \Delta_2^{VA} \Pi_{\rho\pi}^{VA} \left (
V^A + V^T \right ) +
\Delta_4^T  \left (\Pi_{\pi\pi}^T +\Pi_{\pi\pi}^A\right )\left ( V^A + V^T \right ) \nonumber  \\
\label{l} &     & + \Delta_4^A \left ( \Pi_{\pi\pi}^A V^T +
\Pi_{\pi\pi}^T V^A + \Pi_{\pi\pi}^T V^T \right )
\end{eqnarray}
\begin{eqnarray}
\label{m} \Delta_1^{VA} & = & \left ( \Delta_1^{VA}
\Pi_{\rho\rho}^T + \Delta_3^A \Pi_{\rho\pi}^{VA}
+ \Delta_3^T \Pi_{\rho\pi}^{VA} \right )\left ( W^A + W^T \right ) \nonumber \\
\label{n} \Delta_2^{VA} & = & ( \Delta_2^{VA}  \Pi_{\rho\rho}^T +
\Delta_4^A \Pi_{\rho\pi}^{VA} + \Delta_4^T \Pi_{\rho \pi}^{VA} )
\left ( W^A + W^T \right ) \nonumber\\
\label{o}
\Delta_3^{VA} & = & \left ( \Delta_1^A \Pi_{\rho
\pi}^{VA}  + \Delta_1^T \Pi_{\rho\pi}^{VA} + \Delta_3^{VA}
\Pi_{\pi\pi}^A  + \Delta_3^{VA} \Pi_{\pi\pi}^T
\right )( V^A + V^T) \nonumber \\
\label{p} \Delta_4^{VA} & = &   \left (  ( \Delta_2^A +
\Delta_2^T)\Pi_{\rho\pi}^{VA}+ \Delta_4^{VA} \Pi_{\pi\pi}^A +
\Delta_4^{VA} \Pi_{\pi\pi}^T \right )( V^A + V^T )
\end{eqnarray}
We solve these equations sector after sector. Equations (\ref{a} -
\ref{d}) lead clearly to:
\begin{eqnarray}
\Delta_1^A     & = & W^A \nonumber \\
\Delta_2^A     & = & 0   \nonumber \\
\Delta_3^A     & = & 0   \nonumber \\
\Delta_4^A     & = & \frac{V^A}{1 - V^A \Pi^A_{\pi\pi}}
\end{eqnarray}
To proceed further it is convenient to define:
\begin{eqnarray}
R^{TA} & = & W^T + W^A \nonumber \\
V^{TA} & = & V^T + V^A \nonumber \\
V^{LA} & = & V^L + V^A
\end{eqnarray}
Equations (\ref{e} - \ref{h}) contain only $\Delta_j^A$ and
$\Delta_j^L$ components. We obtain:
\begin{eqnarray}
\Delta_1^L     & = &
\frac{W^L + R^{LA}\Pi_{\rho\rho}^L W^A}{1-R^{LA}\Pi^L_{\rho\rho}}\nonumber \\
\Delta_2^L     & = & 0 \nonumber \\
\Delta_3^L     & = & 0 \nonumber \\
\Delta_4^L     & = & \frac{V^L + V^{LA}\Pi_{\pi\pi}^L V^A} {\left
(1 - V^A \Pi_{\pi\pi}^A\right ) \left ( 1 - V^{LA}(\Pi_{\pi\pi}^L
+\Pi_{\pi\pi}^A) \right )}.
\end{eqnarray}
$\Delta^T$ and $\Delta^{VA}$ components mix among themselves but
always in pairs:
\begin{eqnarray}
\Delta_1^T  & \longleftrightarrow & \Delta_3^{VA}\nonumber \\
\Delta_2^T  & \longleftrightarrow & \Delta_4^{VA}\nonumber \\
\Delta_3^T  & \longleftrightarrow & \Delta_1^{VA}\nonumber \\
\Delta_4^T  & \longleftrightarrow & \Delta_2^{VA} .\end{eqnarray}
We derive:
\begin{eqnarray}
\Delta_1^T     & = & \frac{\left [ 1 - V^{TA} (\Pi_{\pi\pi}^A +
\Pi_{\pi\pi}^T)\right ] \left [ W^T +  W^A W^{TA}
\Pi^T_{\rho\rho}\right ] }
{\left[1- V^{TA}(\Pi_{\pi\pi}^A +\Pi_{\pi\pi}^T ) \right] \left[1
- R^{TA} \Pi_{\rho\rho}^T\right] - R^{TA}
V^{TA}(\Pi_{\rho\pi}^{VA})^2}
\nonumber\\
 & & 
\!\!\!\!\!\!\!\!\!\!\!\!\!\!\!\!\!\!\!\!\!\!\!\!
+\frac{ W^A R^{TA}V^{TA}(\Pi_{\rho\pi}^{VA})^2}
{\left[1- V^{TA}(\Pi_{\pi\pi}^A +\Pi_{\pi\pi}^T ) \right] \left[1
- R^{TA} \Pi_{\rho\rho}^T\right] - R^{TA}
V^{TA}(\Pi_{\rho\pi}^{VA})^2}\\
\Delta_2^T     & = & 0   \nonumber \\
\Delta_3^T     & = & 0  \nonumber \\
\Delta_4^T     & = & \frac{(1-R^{TA}\Pi^T_{\rho\rho}) \left (
V^T +
\Delta_4^A( V^{TA}\Pi^T_{\pi\pi}+V^T\Pi^A_{\pi\pi}) \right)}{\left [ 1 -
V^{TA}\left (\Pi^A_{\pi\pi}+\Pi^T_{\pi\pi}\right ) \right ] \left
[1 - R^{TA}\Pi_{\rho\rho}^T \right ]
-V^{TA}R^{TA}(\Pi_{\rho\pi}^{VA})^2 }  \nonumber \\
& & 
\!\!\!\!\!\!\!\!\!\!\!\!\!\!\!\!\!\!\!\!\!\!\!\!
+\frac{
\Delta_4^A V^{TA}R^{TA} (\Pi_{\rho\pi}^{VA})^2}{\left [ 1 -
V^{TA}\left (\Pi^A_{\pi\pi}+\Pi^T_{\pi\pi}\right ) \right ] \left
[1 - R^{TA}\Pi_{\rho\rho}^T \right ]
-V^{TA}R^{TA}(\Pi_{\rho\pi}^{VA})^2}  
\end{eqnarray}
\begin{eqnarray}
\Delta_1^{VA}  & = &  0   \nonumber  \\
\Delta_2^{VA}  & = &  
\frac{W^{AT}\Pi_{\rho\pi}^{VA}
 ( V^T +
(1 - \Pi_{\pi\pi} V^A)\Delta_4^A)}
{1  - W^{AT}(\Pi_{\rho\rho}^T +
V^{AT}(\Pi_{\rho\pi}^{VA})^2)+
V^{AT}(\Pi_{\pi\pi}^A + \Pi_{\pi\pi}^T)(\Pi_{\rho\rho}^T W^{AT}-1)}  \\\Delta_3^{VA}  & = & 
\frac{V^{AT}W^{AT}\Pi_{\rho\pi}^{VA}}
{1  - W^{AT}(\Pi_{\rho\rho}^T +
V^{AT}(\Pi_{\rho\pi}^{VA})^2 )+
V^{AT}(\Pi_{\pi\pi}^A + \Pi_{\pi\pi}^T)(\Pi_{\rho\rho}^T W^{AT}-1)} \\
\Delta_4^{VA}  &  = &  0 \nonumber
\end{eqnarray}

\textbf{\large Acknowledgments}

The work was supported by KBN grant 105/E-344/SPB/ICARUS/P-03/DZ 211/2003-2005. 
The
authors thank anonymous referee for questions, remarks and
suggestions which led to substantial improvement of the paper.

\end{document}